\documentstyle[art12]{article}
\begin{document}
\begin{titlepage}
\begin{flushright}
DTP/95/50\\
FERMILAB-CONF-95-168-T\\
\end{flushright}
\vspace{1cm}
\begin{center}
{\Large\bf Physics with jets in $p\bar p$ collisions}\\
\vspace{1cm}

{\large
W.~T.~Giele}\\
\vspace{0.4cm}
{\it
Fermi National Accelerator Laboratory, P.~O.~Box 500,\\
Batavia, IL 60510, U.S.A.} \\
\vspace{0.4cm}
and \\
\vspace{0.4cm}
{\large
E.~W.~N.~ Glover}\\
\vspace{0.4cm}
{\it
Physics Department, University of Durham,\\ Durham DH1~3LE, England}
\\
\vspace{0.4cm}
{\large \today}
\vspace{0.4cm}
\end{center}

\begin{abstract}

We discuss the possibilities for extracting information on the parton
density functions and the strong coupling constant from one- and
two-jet events at the Fermilab TEVATRON.  First we study the
inclusive
two-jet triply differential cross section
$d^3\sigma/dE_Td\eta_1d\eta_2$.  Different $\eta_1$ and $\eta_2$
pseudorapidity regions are directly related to the parton momentum
fractions at leading order and the shape of the triply differential
distribution at fixed transverse energy $E_T$ is a particularly
powerful tool for constraining the parton distributions at small to
moderate $x$ values.  Second, we consider the one-jet inclusive
transverse energy distribution where there is impressive agreement
between theory and experiment over a wide range of transverse energy.
By equating the next-to-leading order theoretical prediction for a
given set of input parton densities with the published CDF data, the
evolution of the strong coupling constant at scale $\mu$ can be
studied between $\mu \sim 50~$GeV and $\mu \sim 400~$GeV.  This
evolution is in agreement with QCD and corresponds to $\alpha_s(M_Z)$
= 0.121 for the MRSA parameterisation.
\end{abstract}

\end{titlepage}

One- and two-jet production in hadron collisions occurs when two
partons from the incident hadrons undergo a hard pointlike
interaction
and scatter at relatively large angles.  The cross section depends on
the non-perturbative probability of finding a particular parton
inside
the parent hadron, the strong coupling constant and the dynamics of
the hard scattering which can be calculated perturbatively.  In many
cases there is excellent qualitative agreement between the observed
jet cross sections and perturbative QCD calculations at
next-to-leading order based on input parton density functions and
strong coupling constant derived from lower energy experiments.  For
example, the single jet inclusive transverse energy distribution is
well described over eight orders of magnitude.  With this qualitative
agreement in mind, we might expect to be able to use the theoretical
description of the hard scattering to extract the input parameters
such as the distribution of partons in the proton from the data.
This
would be particularly interesting since gluon scattering plays a very
important role in two jet production, and it may be possible to probe
the gluon density in a more direct way than is possible in deeply
inelastic scattering or in Drell-Yan processes.  In this talk, we
briefly discuss how such determinations might be attempted at the
TEVATRON using the inclusive two-jet cross section and the single-jet
inclusive transverse energy distribution.

In comparing theory with experiment, we use the ${\cal
O}(\alpha_s^3)$
Monte Carlo program {\tt JETRAD} for one, two and three jet
production
based on the one-loop $2 \to 2$ and the tree level $2 \to 3$ parton
scattering amplitudes \cite{ES} described in ref.~\cite{GGK2jet}.
This program uses the techniques of refs.~\cite{GG,GGK} to cancel the
infrared and ultraviolet singularities thereby rendering the $2\to 2$
and $2\to 3$ parton processes finite and amenable to numerical
computation.  The parton four momenta are then passed through the
parton level equivalent of the standard `Snowmass' cone
algorithm~\cite{Snowmass} with $\Delta R = 0.7$ to determine the one,
two and three jet cross sections according to the experimental cuts.
Throughout, we shall evaluate the cross section at a renormalisation
and factorisation scale $\mu = E_{T1}$, where $E_{T1}$ is the
transverse energy of the hardest jet in the event.

The inclusive two-jet cross section can be described in terms of
variables most suited to the geometry of the detector; the transverse
energy of the leading jet, $E_T = E_{T1}$, and the pseudorapidities
of
the two leading jets, $\eta_1$ and $\eta_2$.  Recently, the D0
collaboration has presented a preliminary measurement of
$d^3\sigma/dE_Td\eta_1 d\eta_2$
\cite{Weerts,D0sign,Geld} as a function of $\eta_1$ and $\eta_2$ at
fixed $E_T$.
At leading order, $\eta_1$ and $\eta_2$ are directly related to the
parton momentum fractions $x_1$, $x_2$, so that this corresponds to a
measurement of $d^2\sigma/dx_1dx_2$.  Of course, beyond leading
order,
the parton momentum fractions are only approximately determined by
the
transverse energies and pseudorapidities of the two leading jets.

For the typical transverse energies probed by CDF and D0, ${\cal
O}(30-50$~GeV), the values of $x$ range between $x_1 \sim 4E_T^2/s
\sim 10^{-3}$, $x_2 \sim 1$ for $\eta_1 \sim \eta_2 \sim 4$ to
$x_1\sim x_2 \sim 2E_T/\sqrt{s} \sim 0.05$ for $\eta_1 \sim \eta_2
\sim 0$.  This covers both the small $x$ region where gluons with
singular behaviour ($xg(x) \sim x^{-\lambda}$ and $\lambda \sim 0.3 -
0.5$) dominate and the intermediate $x$ region where the momentum sum
rule ensures that the less singular gluon distributions carry a
larger
fraction of the momentum.  This is illustrated in Fig.~1 where
$xg(x)$
is shown for the MRSA parameterisation \cite{mrsa} corresponding to
$\lambda \sim 0.3$ as favoured by the HERA data.  To guide the eye,
we
also show the older pre-HERA MRSD0 and MRSD-
\cite{mrsd} distributions with a flat ($\lambda \sim 0$) and singular
($\lambda\sim 0.5$) behaviour respectively.

\begin{figure}\vspace{8cm}
\includegraphics{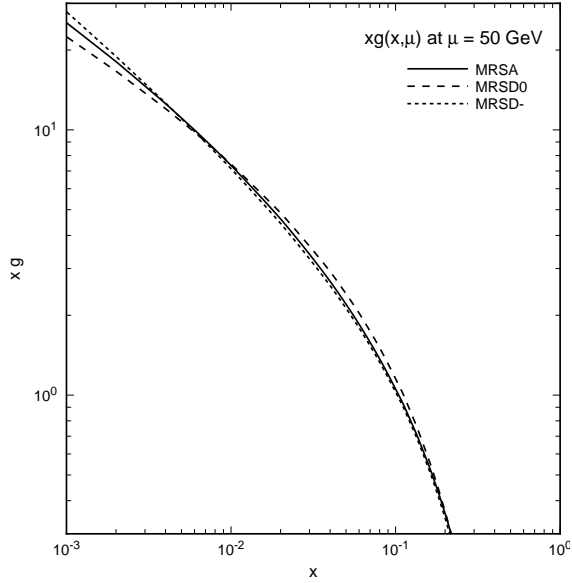}
\caption[]{The momentum fraction $xg(x,\mu)$ carried by the gluon

for $\mu = 50$~GeV for the MRSA, MRSD0 and MRSD- parton densities.}
\end{figure}

Both the CDF \cite{CDFssos} and D0 \cite{Weerts,D0sign,Geld}
collaborations have focused on particular slices of the triply
differential distribution.  D0 study the signed pseudorapidity
distribution which amounts to taking two strips of the
$\eta_1-\eta_2$
plane for a fixed transverse energy interval and combining them in
reverse directions.  The pseudorapidity of the leading jet is
constrained to lie in the range $|\eta_1|_{\rm min} < |\eta_1| <
|\eta_1|_{\rm max}$ and the distribution is plotted as a function of
$|\eta_2| {\rm sign}(\eta_1 \eta_2)$,

$$
\hskip -10pt\frac{d\sigma}{d|\eta_2|{\rm sign}(\eta_1 \eta_2)}
\equiv {1\over\Delta E_T} \int^{E_{T{\rm max}}}_{E_{T{\rm min}}}
dE_T\,
{1\over 2\Delta \eta_1}
\left ( \int^{|\eta_1|_{\rm max}}_{|\eta_1|_{\rm min}} d\eta_1
 \frac{d^3\sigma}{dE_Td\eta_1d\eta_2} -
\int^{-|\eta_1|_{\rm min}}_{-|\eta_1|_{\rm max}} d\eta_1
\frac{d^3\sigma}{dE_Td\eta_1d\eta_2} \right ),
$$

where ${\rm sign}(\eta_1 \eta_2) = -1$ if $\eta_1$ and $\eta_2$
have opposite sign and +1 if they have the same sign.  Positive
values
of $|\eta_2| {\rm sign}(\eta_1 \eta_2)$ correspond to same-side dijet
events, while negative values are associated with opposite-side
events.  Fig.~2(a) shows the preliminary data for 45~GeV $< E_T < $
55~GeV and $0.0 < |\eta_1| < 0.5$.  In the small $|\eta_2|$ region
this is sensitive to $x \sim 0.05$.  The next-to-leading order QCD
predictions \cite{GGK3D} are also shown.  Although the errors are
large and the overall normalisation is uncertain, there is a slight
preference for the MRSD0 parameterisation indicating that perhaps
more
gluons are needed in the intermediate $x$ range than suggested by the
low $x$ data from HERA combined with the momentum sum rule.

\begin{figure}\vspace{8cm}
\includegraphics{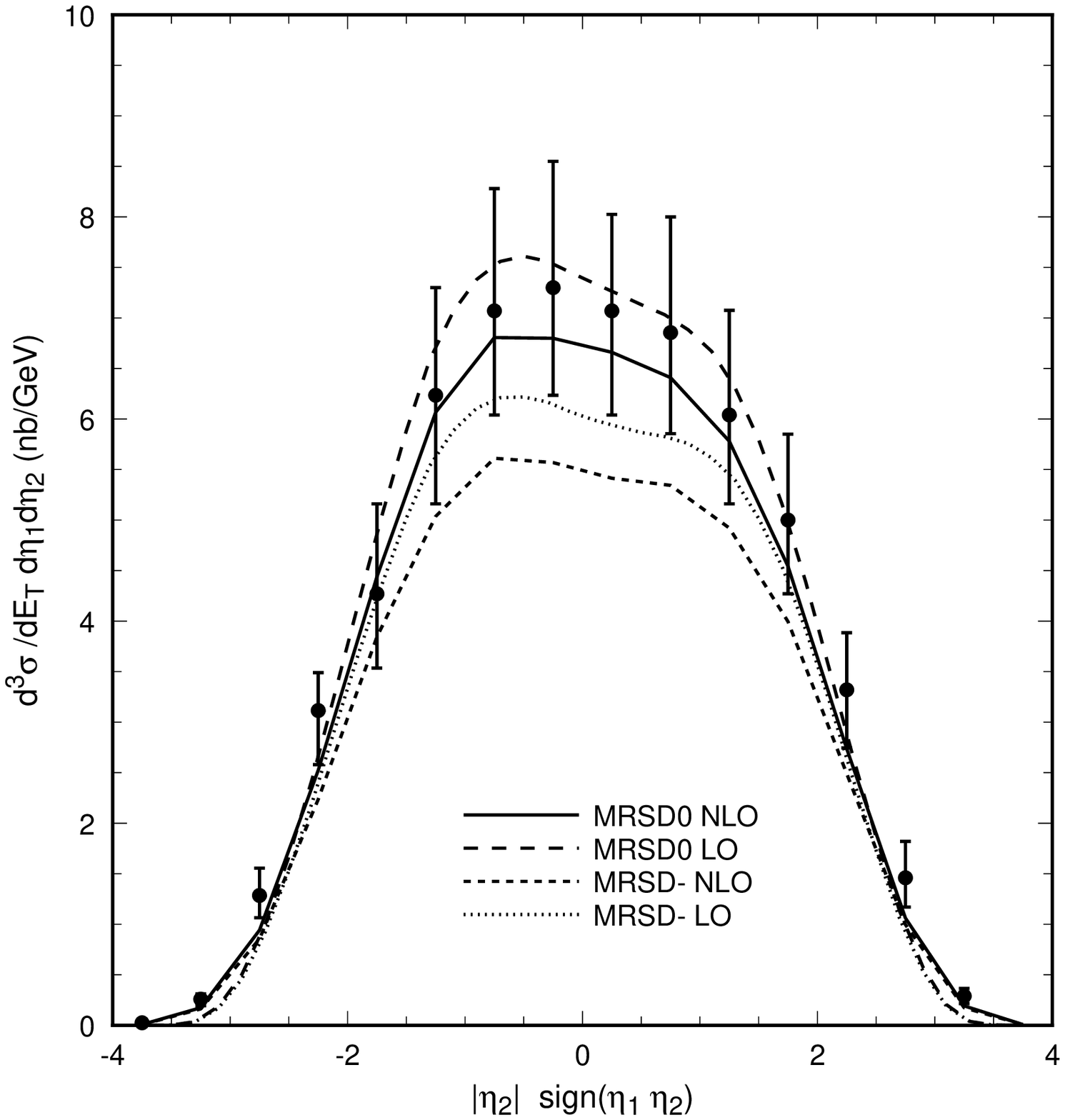}
\includegraphics{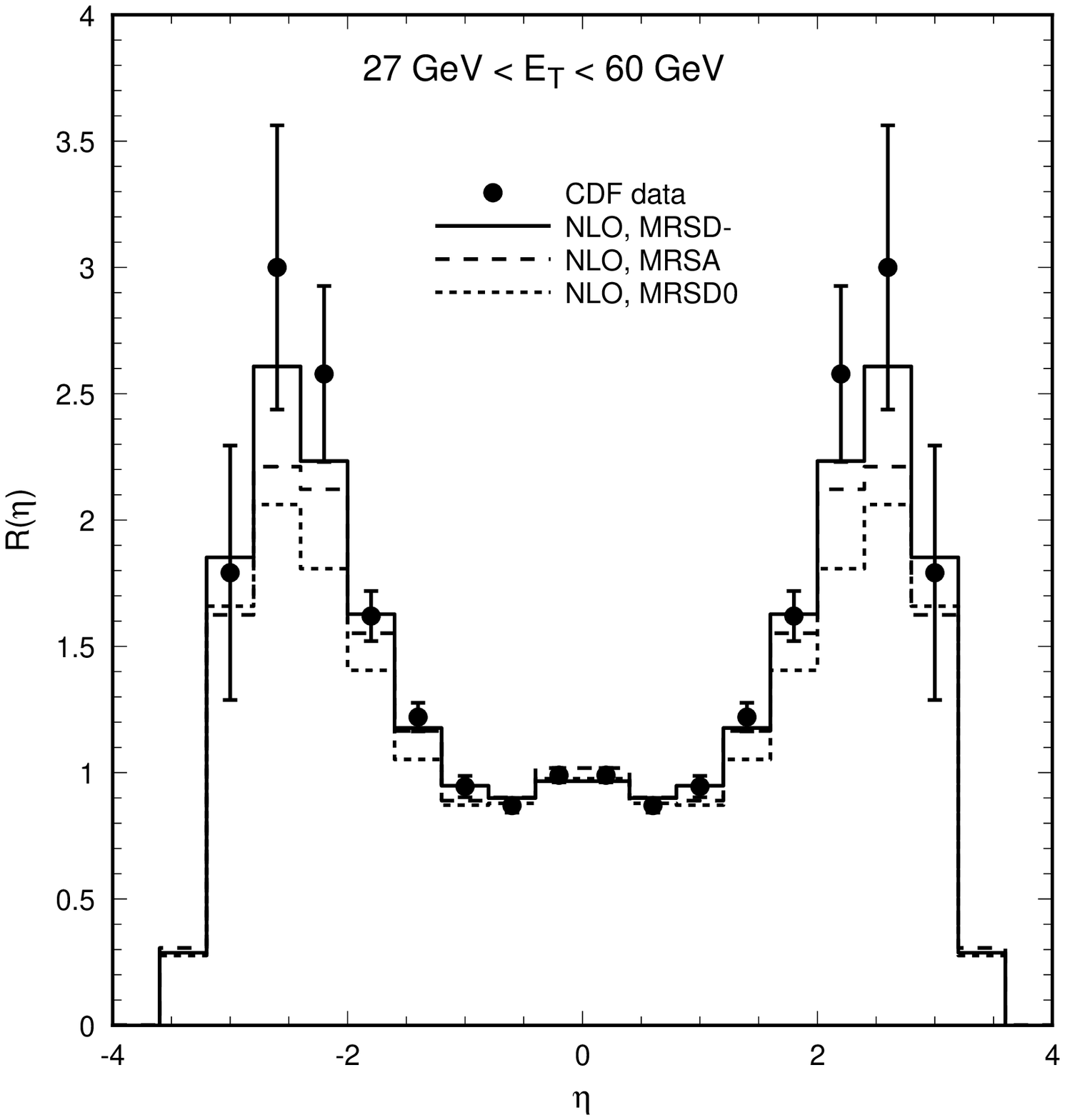}
\caption[]{The next-to-leading order (NLO)

predictions for (a) the signed pseudorapidity distribution for 45 GeV
$< E_T <$ 55 GeV, 0.0 $< |\eta_1| <$ 0.5 and (b) the SS/OS ratio
$R(\eta)$ for 27 GeV $<E_T<$ 60 GeV for the MRSD- (solid) and MRSD0
(dashed) parton distributions. The preliminary experimental results
from \cite{Weerts} and \cite{CDFssos} are also plotted. }
\end{figure}

The interpretation of D0 signed distribution measurement will
ultimately hinge on the absolute normalisation of the cross section.
To circumvent this problem the CDF collaboration has considered a
ratio, that of same-side (SS) to opposite-side (OS) cross
sections~\cite{CDFssos}.  For the same-side cross section, both jets
have roughly the same pseudorapidity, while in the opposite-side
cross
section the jets are required to have roughly equal, but opposite
pseudorapidities,
\begin{eqnarray}
\sigma_{SS}(\eta)\Big \rfloor_{E_{T{\rm min}}<E_T<E_{T{\rm max}}} &=&
\int_{\eta-\Delta \eta}^{\eta+\Delta \eta} d\eta_1
\int_{\eta-\Delta \eta}^{\eta+\Delta \eta} d\eta_2
\int_{E_{T{\rm min}}}^{E_{T{\rm max}}} dE_T
\frac{d^3\sigma}{ dE_Td\eta_1 d\eta_2}, \nonumber \\
\sigma_{OS}(\eta)\Big \rfloor_{E_{T{\rm min}}<E_T<E_{T{\rm max}}} &=&
\int_{\eta-\Delta \eta}^{\eta+\Delta \eta} d\eta_1
\int_{-\eta-\Delta \eta}^{-\eta+\Delta \eta} d\eta_2
\int_{E_{T{\rm min}}}^{E_{T{\rm max}}} dE_T
\frac{d^3\sigma}{ dE_Td\eta_1 d\eta_2 }.\nonumber
\end{eqnarray}
{}From these cross sections we form the SS/OS ratio,

$$
R_{SS/OS}(\eta)\Big \rfloor_{E_{T{\rm min}}<E_T<E_{T{\rm max}}}=\frac
{\sigma_{SS}(\eta)\Big \rfloor_{E_{T{\rm min}}<E_T<E_{T{\rm max}}}}
{\sigma_{OS}(\eta)\Big \rfloor_{E_{T{\rm min}}<E_T<E_{T{\rm max}}}},
$$

with the advantage that a large part of the experimental and
theoretical uncertainties cancel. This ratio loses information on the
parton densities in the central region where $\eta_1\sim \eta_2
\sim 0$.  However, at larger pseudorapidities

it is sensitive to smaller $x$ values $x\sim 4E_T^2/s$.  Fig.~2(b)
shows the preliminary data for 27~GeV $< E_T < $ 60~GeV.  At $\eta
\sim 2.6$, this probes $x \sim 0.001$.  The next-to-leading order QCD
predictions \cite{GGKssos} for which the renormalisation scale
dependence is small are also shown.  Although the experimental errors
are large, there is now a slight preference for the MRSD-
parameterisation indicating that perhaps more gluons are needed in
the
low $x$ range than currently preferred by HERA.

\begin{figure}\vspace{8cm}
\includegraphics{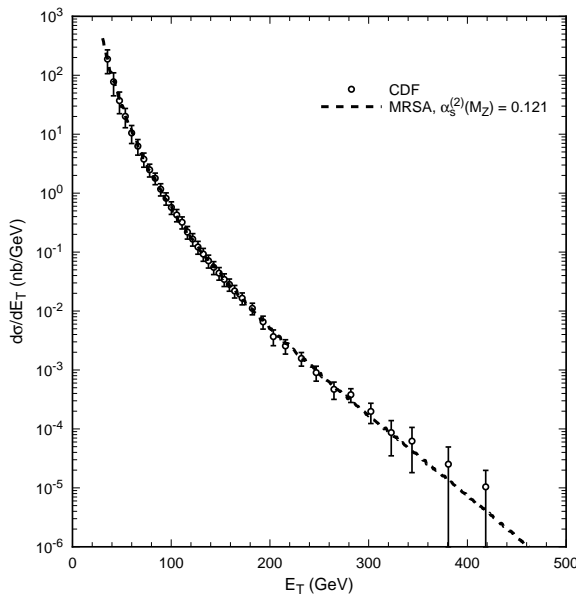}
\caption[]{The next-to-leading order prediction for the single jet

inclusive transverse energy distribution for the MRSA parton
distributions and $\alpha_s^{(2)}(M_Z) = 0.121$.  The CDF data from
\cite{CDFsji} is also shown.}
\end{figure}

We now consider what information on the strong coupling constant
$\alpha_s$ can be gleaned from the single jet transverse energy
distribution.  Up to ${\cal O}(\alpha_s^3)$, this is given by,
\begin{equation}
\frac{d\sigma}{dE_T} = \left (\frac{\alpha_s(\mu)}{2\pi}\right )^2  A
+ \left (\frac{\alpha_s(\mu)}{2\pi}\right )^3\left [ B +
2b_0\log\left(
\frac{\mu^2}{E_T^2}\right) \right ],
\end{equation}
where $\mu$ is the renormalisation scale, $b_0 = (33-2n_f)/6$ and the
next-to-leading order coefficient $B$ has been known for some time
\cite{EKS1,AGCG}.  Fig.~3 shows the excellent agreement between
next-to-leading order QCD and the published CDF data \cite{CDFsji}
over eight orders of magnitude.  We might therefore hope to use this
agreement to extract information on the strong coupling.  However,
both $A$ and $B$ depend on the factorisation scale $\mu_F$ and the
input parton densities (which themselves implicitly depend on
$\alpha_s$).  These dependences will ultimately form part of the
theoretical error in measuring $\alpha_s$ along with the usual
renormalisation scale uncertainty.  For a first look at what can be
learned, we fix $\mu_F = E_T$ and use the MRSA parameterisation of
the
parton densities, ignoring the value of $\alpha_s$ it was derived
from.  We also fix the renormalisation scale to be the transverse
energy of the jet, $\mu=E_T$ and extract a value of $\alpha_s(E_T)$
by
solving Eq.~1 for each of the 38 data points, yielding in principle
38
separate measurements of $\alpha_s$.\footnote{A more careful analysis
would require the experimental error to be resolved into a bin-by-bin
error and a common systematic error.}  This is shown in Fig.~4a along
with the value of $\alpha_s(E_T)$ obtained from $\alpha_s^{(2)}(M_Z)
=
0.121$ using the renormalisation group evolution.  The running of the
strong coupling is clear and is qualitatively in agreement with
expectations.

We can take this one step further by using the renormalisation group
equations to evolve the extracted value of $\alpha_s(E_T)$ back to
$\mu= M_Z$ for each data point, yielding 38 estimates of
$\alpha_s(M_Z)$ as shown in Fig.~4b.  The results are essentially
independent of the $E_T$ at which $\alpha_s(M_Z)$ was extracted and
the error weighted average is $<\alpha_s(M_Z)> = 0.121$.  This can be
compared with the input value used in the parton distribution
evolution of $\alpha_s(M_Z)=0.111$.  As mentioned earlier, the value
of $\alpha_s$ extracted in this way also depends on the choice of
input parton densities.  For the MRSD0 parameterisation, the same
analysis gives $<\alpha_s(M_Z)> = 0.118$ while for the MRSD- set we
find $<\alpha_s(M_Z)> = 0.123$.

Clearly there is more work to be done in understanding both the
theoretical and experimental error.  However, the rewards of
observing
the evolution of the strong coupling over nearly an order of
magnitude
of scale in a single experiment would be considerable, particularly
when we bear in mind that the current TEVATRON run should obtain 25
times more data than that analysed here.  Similarly, one would expect
that with the increase in statistics of the dijet data, it might be
possible to track the high $Q^2$ evolution by varying the transverse
energy of the jet as well as providing a significant constraint on
the
parton distributions.\\

\begin{figure}\vspace{8cm}
\includegraphics{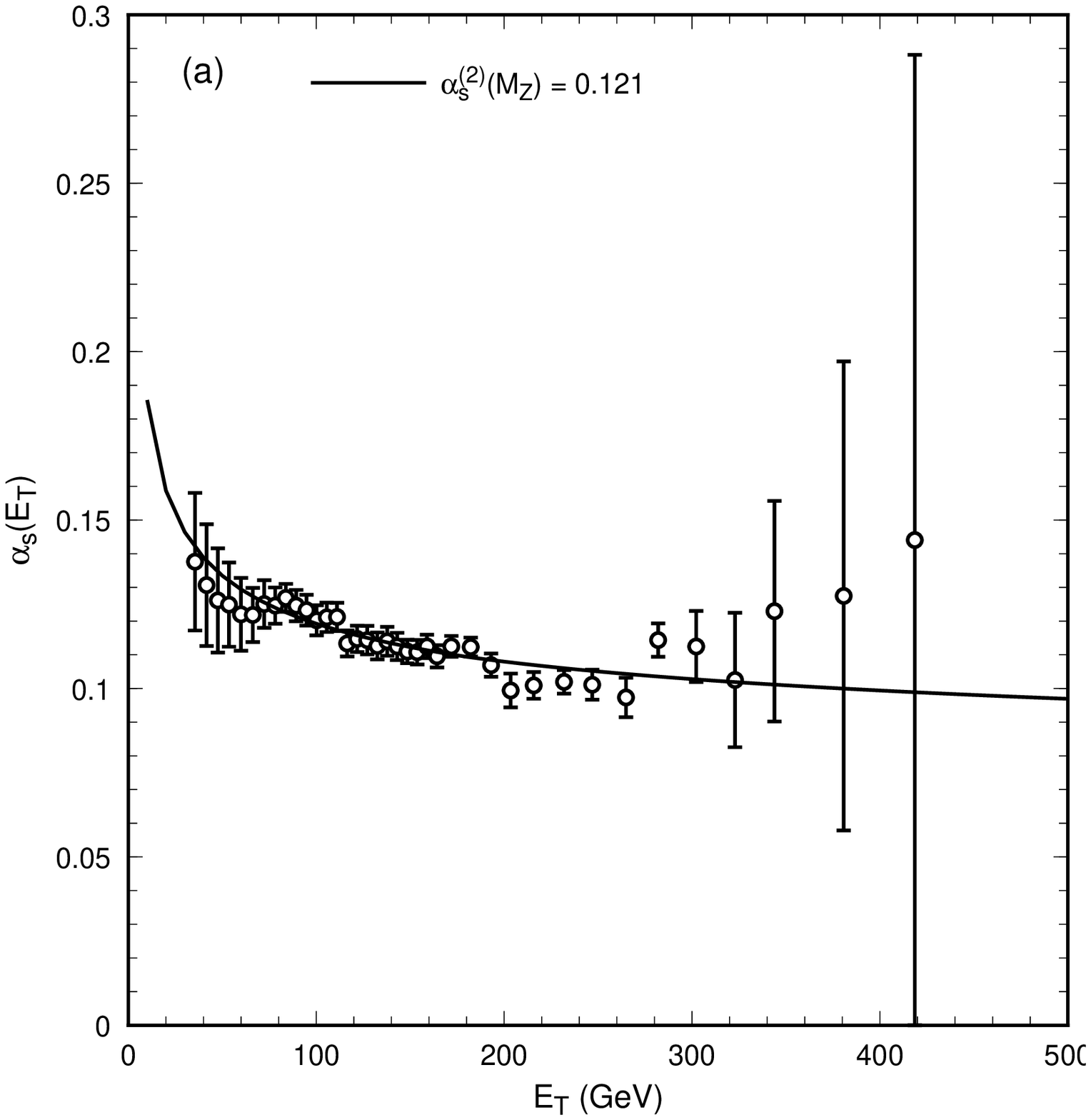}
\includegraphics{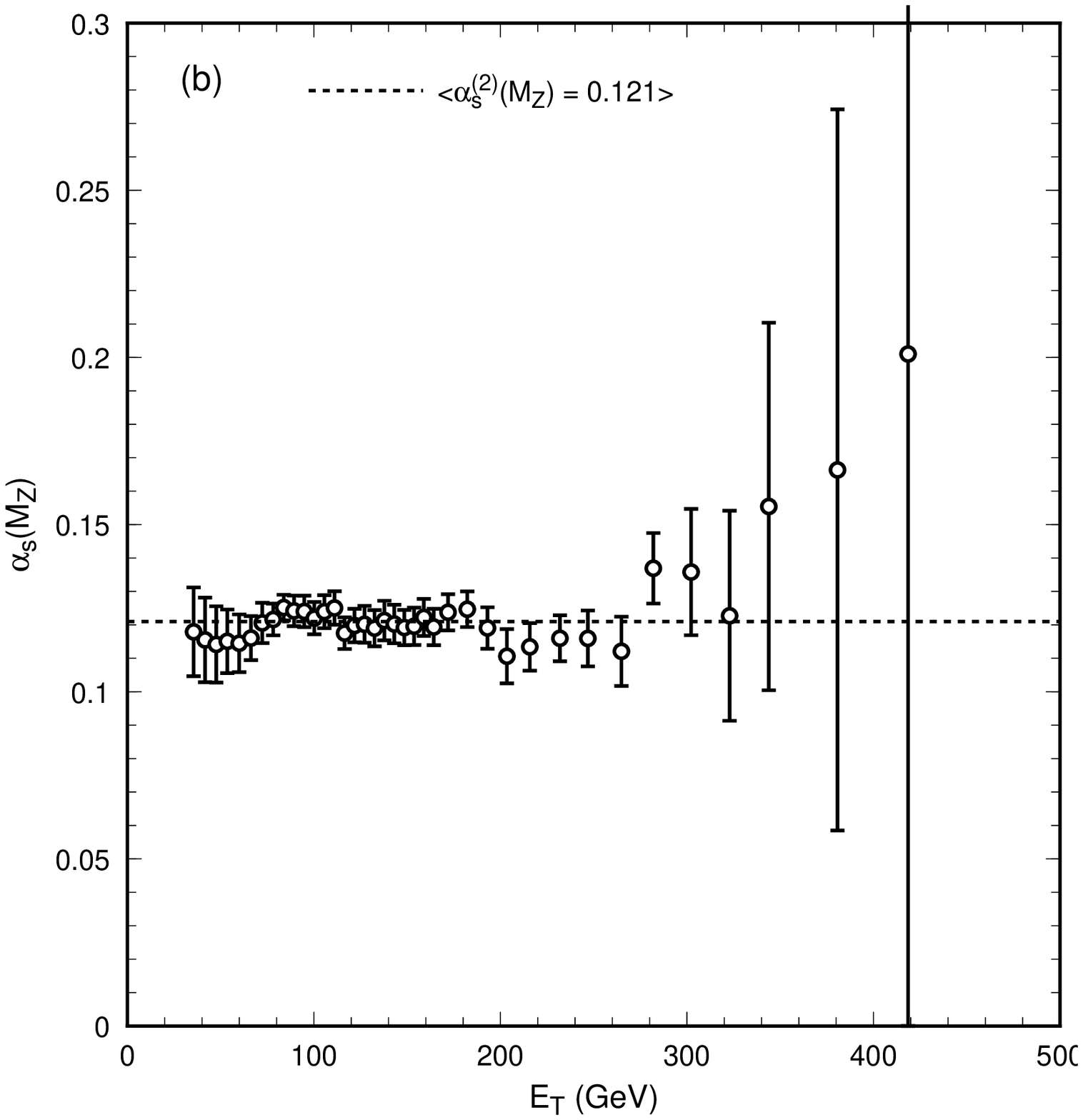}
\caption[]{(a) The value of $\alpha_s(E_T)$ extracted using the
published

CDF data \cite{CDFsji} and Eq.~1 compared to the next-to-leading
order
evolution with $\alpha_s(M_Z)=0.121$ (solid) and (b) the value of
$\alpha_s(M_Z)$ obtained when evolving from $\mu = E_T$ to
$\mu=M_Z$.}
\end{figure}

\newpage
\section*{Acknowledgements}
We thank David Kosower for many valuable insights
throughout a stimulating collaboration.  EWNG gratefully acknowledges
the
financial support of the EC Human Capital and Mobility Network
contract ERBCHRXCT930357.

\end{document}